\documentclass[pre,reprint,nofootinbib,superscriptaddress]{revtex4-2}
\pdfoutput=1

\usepackage{amssymb, amsmath, amsmath, amsfonts, bm, tabularx,graphicx,float}
\usepackage[colorlinks,linkcolor=blue,urlcolor=blue,citecolor=blue]{hyperref}
\usepackage{physics}

\usepackage[english]{babel}

\usepackage{xcolor}
\usepackage{xspace}
\usepackage{ifthen}

\begin{document}

\title{On the strangeness of quantum probabilities}

\author{Marcello {Poletti}}
\email{epomops@gmail.com}
\affiliation{San Giovanni Bianco, Italy}

 \begin{abstract}
Here we continue with the ideas expressed in "On the strangeness of quantum mechanics" \cite{Poletti} aiming to demonstrate more concretely how this philosophical outlook might be used as a key for resolving the measurement problem. We will address in detail the problem of determining how the concept of undecidability leads to substantial changes to classical theory of probability by showing how such changes produce a theory that coincides with the principles underlying quantum mechanics.
\end{abstract}

\maketitle

\section{Introduction}

The work that serves as the starting point for this paper \cite{Poletti} suggests that the extravagances of quantum mechanics, in particular the violation of Bell's inequalities, are connected to the problem of logical undecidability
.
The term \textit{undecidable} refers to a proposition that, within formal system $S$, cannot be shown to be true or false.

At the beginning of the twentieth century, Hilbert \cite{Hilbert} explained the need to formally demonstrate of the completeness of the foundations of mathematics, that is, to demonstrate that every true mathematical proposition can be proven.

In 1931, Gödel \cite{Godel,Lolli} disregarded Hilbert's program by demonstrating how, on the contrary, for any coherent formal system that is sufficiently expressive to include arithmetic, it is possible to generate true propositions that cannot be proven within that system.

Around 1910, Charles Sanders Peirce \cite{Peirce} defined a logical system, expressly proposing a third alternative value to truth and falsehood, which he defined as \textit{limit}. This approach -- ingenuous, in light of the subsequent results -- directly undermines the law of excluded middle by violating the fundamental true/false duality.
The theorem of incompleteness may seem to suggest that Gödel was doing the same, but this is a gross error - according to Gödel, \textit{undecidable} is not an alternative to true or false.

We define as \textit{manifestly true} in $S$ a proposition $p$ which can be proven in $S$. A proposition whose negation, instead, can be proven in $S$ is then said to be \textit{manifestly false}.

We shall refer to true and false with the swash-serif symbols $\mathcal{T}$ and $\mathcal{F}$. We shall refer to manifestly true and false with the straight-serif symbols $\mathsf{T}$ and $\mathsf{F}$. Undecidability will be indicated with the symbol $\mathsf{U}$.

To rephrase, $\mathsf{U}$ is an alternative condition to $\mathsf{TF}$, not to $\mathcal{TF}$. Thus we create two domains of propositions, the binary system $\mathcal{TF}$ subject to the law of excluded middle, and the ternary $\mathsf{TFU}$ system, not subject to that law and connected to the concept of demonstrability.

Both these domains, $\mathcal{TF}$ and $\mathsf{TFU}$, fall within classical logic. Reasoning about the implications of the $\mathsf{TFU}$ domain has nothing to do with the elaboration of \textit{another logic}, and in this the work proposed here diverges immediately from the classical works called many-valued logics \cite{ManyValuedLogic}, as well from what is known as \textit{quantum logic}, understood as logic in any way alternative to classical logic.

A $\mathsf{TFU}$ perspective has obvious repercussions in various contexts. Boolean logic, for example, handles the $\mathcal{T}$ and $\mathcal{F}$ values very clearly, and thus, within a given system $S$ 
there are propositions for which algebra cannot be used, propositions for which it is not concretely possible to associate any Boolean value $\{0,1\}$.

It is possible, however, to conceive of a $\mathsf{TFU}$ Boolean algebra that handles the concept of demonstrability  and that produces an \textit{alternative Boolean algebra} within the confines of classical logic. An algebra of this kind will appear paradoxical whenever we make the error of confusing truth with demonstrability.

A second excellent example is offered by the theory of probability. 
\textit{Probability} is a borderline concept, straddling mathematical theory and the pragmatics of measurement. It is about the frequencies with which certain events, a priori unknown, \textit{occur} and this raises an issue: is this \textit{occurrence} a manifestation of the truth ($\mathsf{T}$) or an ontological state of being true ($\mathcal{T}$)? What is probability? A relative measurement of the states in which a certain proposition is true, or a relative measurement of the states in which it is manifestly true?

It is possible to develop (at least) two theories of probability - one from a $\mathcal{TF}$ perspective and one from a $\mathsf{TFU}$ perspective. Ultimately, the thesis of this paper is that the former coincides with the classical theory of probability and the latter with the foundational principles of quantum mechanics.

In essence, we argue here that the paradoxes of quantum mechanics can be traced to a fundamental confusion between $\mathcal{TF}$ and $\mathsf{TF}$. The overlapping of states such as $\ket{True}+\ket{False}$ ("both true and false" or "neither true nor false") violate the excluded middle and lead to the search for logical "alternatives" (that violate the excluded middle as in Pierce, or that have truth continuums, as in fuzzy logic \cite{Zadeh} or others, etc.) or even result in a search for extraordinarily costly models that in some way restore \textit{logicality} (many worlds \cite{Everett,Vaidman}, many minds \cite{Albert}).

These paradoxes immediately dissolve when the overlapping status is written as $\ket{Manifestly \,True}+\ket{Manifestly \,False}$(neither $\mathsf{T}$ nor $\mathsf{F}$, undecidable), which is not illogical at all and does not require any violation of logical principles; in particular, it does not violate the law of excluded middle.

The theorem of incompleteness has often been associated with the strangeness of quantum mechanics, not infrequently in terms that verge on metaphysics. Without detracting anything from those arguments, the point of departure here is that this theorem also has very tangible implications and that such implications may be the key for bringing the apparent illogicality of QM back into the completely logical domain.

In \cite{Poletti}, violation of the Wigner-d'Espagnat inequality \cite{Bell} is used as a litmus test for the reasoning applied:

\begin{equation*}
	(A,B)\cup(\neg{B},C)\supseteq(A,C)
\end{equation*}

Things that are \textit{A e B}, together with things that are \textit{not B and C}, including things that are \textit{A and C}.
This same relationship is rewritten in $\mathsf{TFU}$ terms: Things that are \textit{manifestly A and B}, are combined with things that are \textit{manifestly not B and C}, including things that are \textit{manifestly A and C}. In this form, the inequality is immediately violated by considering things that are manifestly A and C and for which the B property is undecidable.

We will proceed with a few preliminary considerations on the possibility of a Boolean $\mathsf{TFU}$ algebra.
We will then formulate the classical theory of probability in a matrix-vector format, to show the connections with QM and to prepare an extension to a $\mathsf{TFU}$ probability theory.
Finally, we will explicitly construct the foundations of such a theory and show how they coincide with the first postulates (excluding Shrödinger's dynamic equation) of quantum mechanics.

\section{Terminology}

\subsection*{Complete States}

Given $n$ propositions $p_1$, $p_2$, etc., we define \textit{complete state} as the conjunction of all $p_i$ or their negations. For example, given $p$ and $q$, the complete states would be the propositions $p\land q$, $p\land\neg q$, $\neg p\land q$, $\neg p\land\neg q$.

The complete states of $n$ propositions are $2^n$.

The conjunction of two distinct complete states is a contradiction.

The disjunction of all $2^n$ complete states is a tautology.

\subsection*{Vectors, Directions, Operators}

A direction is a vector space, i.e., a sub-space with a dimension of $1$ will be called a \textit{state} and will henceforth be indicated, using Dirac notation, with a capital letter, for example $\ket{S}$, $\ket{P}$, $\ket{Q}$,... 

Generic vectors will be indicated with lower-case letters, for example $\ket{s}$, $\ket{p}$, $\ket{q}$...

Versors will be indicated in bold-face type, for example $\ket{\bm{s}}$, $\ket{\bm p}$, $\ket{\bm q}$...

Linear operators will be indicated with a bold-faced capital letter, for example $\bm P$, $\bm Q$, ...
Since a linear operator maps vectors of the same direction on the same direction, the symbol $\bm P\ket{S}$ will be used to indicate the common direction for each vector $k\bm P\ket{s}$.

The symbol $\braket{P}{S}$ will be used for the cosine of the angle subtended by directions $\ket P$ and $\ket S$, i.e., $\braket{P}{S}=\braket{\bm p}{\bm s}$.\footnote{This leads to an ambiguity about the sign which is however irrelevant}

\section{TFU Boolean algebra}

First we will take a look at the basic properties of a possible $\mathsf{TFU}$ Boolean algebra that focuses on the demonstrability of the propositions rather than the absolute concepts of true or false.

The negation of a proposition  $\mathsf{T}$ is $\mathsf{F}$, and vice versa. The negation of a proposition $\mathsf{U}$ is $\mathsf{U}$.

\begin{equation*}
	\begin{cases}
		\neg \mathsf{T}=\mathsf{F} \\
		\neg \mathsf{F}=\mathsf{T} \\
		\neg \mathsf{U}=\mathsf{U}
	\end{cases}
\end{equation*}

For the conjunction of two propositions $p$ and $q$ a truth table is applied:

\setlength\extrarowheight{4pt}
\begin{table}[h!]
	\begin{tabularx}{.3\textwidth}{|c|>{\centering\arraybackslash}c|>{\centering\arraybackslash}X|>{\centering\arraybackslash}X|>{\centering\arraybackslash}X|}
		\multicolumn{3}{c}{} & \multicolumn{1}{c}{$p$} & \multicolumn{1}{c}{} \\ 
		\multicolumn{1}{c}{}  & \multicolumn{1}{c}{$\land$} & \multicolumn{1}{c}{$\bm{\mathsf{T}}$} & \multicolumn{1}{c}{$\bm{\mathsf{F}}$} & \multicolumn{1}{c}{$\bm{\mathsf{U}}$} \\ \cline{3-5}
		\multicolumn{1}{c}{} & $\bm{\mathsf{T}}$ & $\mathsf{T}$ & $\mathsf{F}$ & $\mathsf{U}$ \\ \cline{3-5}
		\multicolumn{1}{c}{$q$} & $\bm{\mathsf{F}}$ & $\mathsf{F}$ & $\mathsf{F}$ & $\mathsf{F}$ \\ \cline{3-5}
		\multicolumn{1}{c}{} & $\bm{\mathsf{U}}$ & $\mathsf{U}$ & $\mathsf{F}$ & $\mathsf{U/F}$ \\ \cline{3-5}
	\end{tabularx}

\end{table}

The conjunction $p\land q$ is manifestly true if and only if their conjoined parts are true.

If one of the conjoined parts is manifestly false, then so is that conjunction.

The conjunction of an undecidable proposition and a manifestly true proposition is undecidable.

The case $\mathsf{U}\land\mathsf{U}$ is critical. The conjunction of two undecidable propositions may in fact be both undecidable and false, in which case $p\implies\neg q$ ($q\implies\neg p$).\footnote{The implication is understood here more in a semantic sense than a material one. The manifest falseness of $\mathsf{U}\land\mathsf{U}$ is directly connected to the presence of the logical nexus $p\implies\neg q$. The set of manifestly false complete states for undecidables $p$ and $q$ completely defines the semantic relationships between $p$ and $q$, in terms of the connections $p\implies q$, $p\implies\neg q$, $p\iff q$...} 

Conversely, when the $\mathsf{TFU}$ values for the complete states are known, it is possible to unambiguously determine the $\mathsf{TFU}$ values of the conjoined parts, in fact, given $n$ propositions $p$, $q$, $r$...

\begin{enumerate}
	\item[I.] $p$ is $\mathsf{F}$ if and only if every complete state that includes $p$ in affirmative form is of type $\mathsf{F}$.
	\item[II.] $p$ is $\mathsf{T}$ if and only if every complete state that includes $p$ in negative form is of type $\mathsf{F}$.
\end{enumerate}

In one direction I. is obvious: if $p$ is manifestly false, every complete state that includes $p$ will in turn be manifestly false.

For the opposite direction, $\lor_{p,q,r,...}$ indicates the disjunction of the complete states of $p$, $q$, $r$.... If every complete state that includes $p$ in affirmative form is of type $\mathsf{F}$, then the disjunction $o$ of all these same states will be type $\mathsf{F}$. But $o=p\land\lor_{q,r...}$ and $\lor_{q,r...}$ is a tautology for which the falseness of $o$ entails the falseness of $p$.

II. derives from the observation that $p$ is $\mathsf{T}$ if and only if $\neg p$ is $\mathsf{F}$.

A $\mathsf{TFU}$ Boolean algebra cannot be the mere syntax of the three symbols $\mathsf{T}$, $\mathsf{F}$, $\mathsf{U}$ or their immediate algebraic representations, because this would sacrifice the concept of a static truth table. The state of propositions $p$, $q$, $r$, ... is not completely given by the respective values $\mathsf{TFU}$, while it may be given by the values $\mathsf{TFU}$ of the complete states, which, in addition to providing values for the conjoined parts, specify the semantic relationships between the propositions. In particular, the assignment:

\begin{equation*}
	p_u\land q_u=\mathsf{F}
\end{equation*}

is the index of the nexus  $p\implies\neg q$.

\section{CTP matrix representation}

The classical theory of probability (CTP) may be expressed in a matrix/vector form which will now be described. The purpose of this representation is to highlight the relationships and differences between CTP and QM, and to prepare an algebraic foundation for the explicit construction of a $\mathsf{TFU}$ theory of probability.

In CTP, the rule of conditional probability applies:

\begin{equation}\label{peq=pqp}
|p\land q|=|p| |q|_p
\end{equation}

Where $|q|_p$ is a new algebraic symbol associated with probability \textit{$q$, known $p$}. 

Once again, the probability value of the conjunction is not the mere syntax of probability values for the parts, but depends on the semantic relationships between the propositions. Conversely, if the probability values of the 4 complete states of $p$ and $q$ are known, then it is possible to determine the probabilities of the conjoined parts. In fact, by combining \ref{peq=pqp} with the fundamental rule

\begin{equation}\label{notp1menop}
	|\neg p|=1-|p| 
\end{equation}

we obtain:

\begin{equation*}
	|p|=|p|(|q|_p+|\neg q|_p)=|p\land q|+|p\land\neg q|
\end{equation*}

More generally, $n$ propositions generate $2^n$ complete states. These states are always separate and exhaustive and cannot be further separated. In this respect, a complete information system in CTP is given by the $2^n$ probability values associate with those conjunctions, more specifically: the probability of a generic proposition $\phi$, decidable when the $p$, $q$, $r$... truth values are known, can be calculated by simply adding together the probabilities associated with the complete states that verify $\phi$. 

The $2^n$ complete states $s_1$, $s_2$...  are associated with the versors $\ket{\bm s_1}$, $\ket{\bm s_2}$... of an orthonormal base of vector space $\mathbb{R}^{2^n}$.

In this way we build the vector:

\begin{equation*}
	\ket{\bm s}=\sum_i{\sqrt{|s_i|}\ket{\bm s_i}}
\end{equation*}

$\ket{\bm s}$ is in turn a versor, whose components in relation to base  $\{s_i\}$ are the roots of the probabilities for the corresponding complete states. 

A generic proposition $p$ is associated with diagonal linear operator  $\bm P$ defined as follows:
\begin{itemize}
	\item $\bm P_{ii}=1$ If $p$ is true in state $s_i$
	\item $\bm P_{ii}=0$ If $p$ is false in state $s_i$
\end{itemize}
Projector $\bm P$ then acts on vector $\ket v$ by annihilating the $\ket v$ components corresponding to the complete states in which $p$ is false and by keeping unchanged the components corresponding to the complete states in which $p$ is true. Thus we have:

\begin{equation*}
	|p|=\|\bm P\ket{\bm s}\|^2
\end{equation*}

And since $\bm P$ is idempotent:

\begin{equation*}
	|p|=\expval{\bm P}{\bm s}
\end{equation*}

We find immediately that the proposition $\neg p$ will be associated with the operator:

\begin{equation*}
	\neg \bm P=\bm I-\bm P
\end{equation*}

And that proposition $p\land q$ will be associated with the operator:

\begin{equation*}
	\bm P\land \bm Q=\bm P\bm Q=\bm Q\bm P
\end{equation*}

In this algebraic form, the probability of conjunction  $p\land q$ is a direct function of the symbols associated with $p$ and $q$ and CTP turns out to be based on the same underlying rules as classsical Boolean algebra.

These rules enable us to derive the algebraic representation of arbitrary combinations of the given propositions, in particular:

\begin{equation*}
	\bm P\lor \bm Q=\bm P+\bm Q-\bm P\bm Q
\end{equation*}

Every tautology is true in every one of the complete states and will thus be associated with identity operator $\bm I$. Every contradiction will be false in every one of the complete states and will be associated with the null operator  $\bm 0$. In particular, we have

\begin{equation*}
	\begin{cases}
		\bm P\land \neg\bm P=\bm 0 \\
		\bm P\lor \neg\bm P=\bm I
	\end{cases}
\end{equation*}

We pose:

\begin{equation*}
	\begin{cases}
		\ket{p_s}=\bm P\ket{\bm s} \\
		\ket{\neg p_s}=(\bm I-\bm P)\ket{\bm s}=\neg\bm P\ket{\bm s}
	\end{cases}
\end{equation*}	

Subscript $s$ makes the statement excessively elaborate, so we will use the simpler $\ket{p}$ in place of $\ket{p_s}$. However, it is important to bear in mind that, unlike the operator  $\bm P$, the direction $\ket{P}$ does not represent proposition $p$ itself, but rather a state related to $\ket{S}$. 

Direction $\ket{S}$ is a linear combination of $\ket{P}$ and $\ket{\neg P}$:

\begin{equation*}
	\begin{cases}
		\ket{\bm s}=(\bm P+\bm I-\bm P)\ket{\bm s}=\alpha\ket{\bm p}+\beta\ket{\neg \bm p} \\
		\alpha^2=|p| \\
		\beta^2=|\neg p| \\
		\alpha^2+\beta^2=1
	\end{cases}
\end{equation*}	

The square of the cosine of the angle between states $\ket{S}$ and $\ket{P}$ is $|p|$:

\begin{equation*}
	\braket{P}{S}^2=\braket{\bm p}{\bm s}^2=\frac{\braket{p}{\bm s}^2}{|p|}=\frac{\expval{\bm P}{\bm s}^2}{|p|}=|p|
\end{equation*}	

Direction $\ket{P}$ may be reconceived, in turn, as a state in which the probability of $p$ is $1$, and indeed we have:

\begin{equation*}
	\expval{\bm P}{\bm p}=1
\end{equation*}		

This is because $\ket{\bm p}$ is the eigenversor of $\bm P$ with eigenvalue $1$.

The probability of $q$ for relative to state $\ket{P}$ will be the probability of \textit{$q$, known $p$}:

\begin{equation*}
	\expval{\bm Q}{\bm p}
	=\frac{\expval{\bm{PQP}}{\bm s}}{|p|}
	=\frac{\expval{\bm{PQ}}{\bm s}}{|p|}
	=\frac{|p\land q|}{|p|}
	=|q|_p
\end{equation*}		

Where $\bm{PQP}=\bm{PQ}$ is guaranteed by the commutativity and idempotence of $\bm P$.

And once again, the cosine squared of the angle between $\ket{P}$ and $\ket{P\land Q}$ gives $|q|_p$

\begin{equation*}
	\braket{P}{P\land Q}^2
	=\frac{\matrixel{\bm{p}}{\bm{PQ}}{\bm s}^2}{|p\land q|}
	=\frac{\matrixel{\bm{s}}{\bm{PQ}}{\bm s}^2}{|p\land q||p|}
	=\frac{|p\land q|}{|p|}
	=|q|_p
\end{equation*}		

Operator $\bm P$ acts, then, by transforming $\ket{\bm s}$ in the new vector $\ket{p}=\bm P\ket{\bm S}$ and the probability of this transition is a function of the angle between the two vectors. A second operator $\bm Q$ then acts on $\ket{p}$ and transforms it into $\bm{QP}\ket{\bm S}$ and $|q|_p$ will be the cosine squared of the angle between these two vectors.
\begin{figure}[H]
	\centering
	\includegraphics[width=0.7\linewidth]{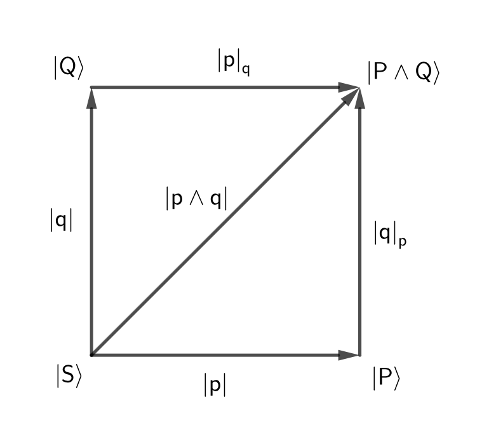}
	\caption[Didascalia]{}
	\label{fig:img1}
\end{figure}

Finally, the cosine squared of the angle between $\ket{P}$ and $\ket{Q}$ gives us:

\begin{equation*}
	\braket{P}{Q}^2
	=\frac{\matrixel{\bm{s}}{\bm{PQ}}{\bm s}^2}{|p||q|}
	=\frac{|p\land q|^2}{|p||q|}
	=\frac{|p||q|_p|q||p|_q}{|p||q|}
	=|p|_q|q|_p
\end{equation*}		

Expressed in this form, the classical theory of probability shows certain affinities with the algebraic mechanisms used in quantum mechanics. The entire amount of information on the semantic relationships between propositions is given a certain direction $\ket{S}$ in an appropriate vector space. The propositions are associated with diagonal linear operators (or all simultaneously diagonalizable by applying a general change of basis). The probability $|p|$ is given by the square of the cosine of angle $\widehat{PS}$ in a form wholly analogous to the Born rule. The evident differences between this algebraic form and QM are the use of the real field instead of a complex one and a wider selection of operators, in QM, including in particular operators that do not commute.

\section{Extension of CTP to TFU}

The classical theory of probability may be based on the concept of measure. In very basic terms:

\begin{equation}\label{ctpdef}
	|p|=\frac{\|\mathcal{T}\|}{\|\mathcal{T}\|+\|\mathcal{F}\|}
\end{equation}		

In other words, the probability of $p$ is defined as the ratio between the measure of states in which $p$ is true and the measure of possible states. 

Considering two propositions $p$ and $q$ we have $4$ possible states $(\mathcal{T},\mathcal{F})\otimes(\mathcal{T},\mathcal{F})$:

\begin{equation*}
	\begin{cases}
		|p|=\frac{\|\mathcal{TT}\|+\|\mathcal{TF}\|}{\|\mathcal{TT}\|+\|\mathcal{TF}\|+\|\mathcal{FT}\|+\|\mathcal{FF}\|}=|p\land q|+|p\land\neg q| \\
		|q|=\frac{\|\mathcal{TT}\|+\|\mathcal{FT}\|}{\|\mathcal{TT}\|+\|\mathcal{TF}\|+\|\mathcal{FT}\|+\|\mathcal{FF}\|}=|p\land q|+|\neg p\land q|
	\end{cases}
\end{equation*}		

Probability \textit{$q$, known $p$} can easily be identified by reducing $|q|$ to only those states in which $p$ is $\mathcal T$:

\begin{equation*}
	|q|_p=\frac{\|\mathcal{TT}\|}{\|\mathcal{TT}\|+\|\mathcal{TF}\|}
\end{equation*}		

Thus, we obtain the rule of conditional probability:

\begin{equation*}
	|p||q|_p
	=\frac{\|\mathcal{TT}\|}{\|\mathcal{TT}\|+\|\mathcal{TF}\|+\|\mathcal{FT}\|+\|\mathcal{FF}\|}
	=|p\land q|
\end{equation*}		

In addition, by inverting the roles of $p$ and $q$ we obtain the fundamental:

\begin{equation*}
	|p||q|_p=|q||p|_q
\end{equation*}		

We now consider an analogous approach from a $\mathsf{TFU}$ perspective.

For the sake of clarity, the $\mathsf{TFU}$ probability value of proposition $p$  will be indicated with $[p]$, using the symbol $[\cdot]$ to distinguish it from the corresponding $|\cdot|$ used in CTP.

\begin{equation}\label{tfudef}
	[p]=\frac{\|\mathsf{T}\|}{\|\mathsf{T}\|+\|\mathsf{F}\|}
\end{equation}		

In other words, $\mathsf{TFU}$ probability $[p]$ is defined as a measure of the states in which $p$ is manifestly true relative to the states in which it manifests.

The symbol $\mathsf{U}$  does not appear, and it gives the fundamental rule \ref{notp1menop}\footnote{This rule is a form of excluded middle, in $\mathsf{TFU}$ probability it states that only probabilities $\mathsf{T}$ and $\mathsf{F}$ are given, not $\mathsf{U}$.}:

\begin{equation*}
	[\neg p]=1-[p]
\end{equation*}

\ref{ctpdef} and \ref{tfudef} are formally identical and are distinguished only by the epistemological interpretation of the symbols. The only tangible difference between the two approaches is the composition of the propositions . Probability $p$ as a function of pairs $(\mathsf{T},\mathsf{F},\mathsf{U})\otimes(\mathsf{T},\mathsf{F},\mathsf{U})$ gives us:

\begin{equation*}
		[p]=\frac{\|\mathsf{TT}\|+\|\mathsf{TF}\|+\|\mathsf{TU}\|}{\|\mathsf{TT}\|+\|\mathsf{TF}\|+\|\mathsf{TU}\|+\|\mathsf{FT}\|+\|\mathsf{FF}\|+\|\mathsf{FU}\|}
\end{equation*}		

And for $q$:

\begin{equation*}
	[q]=\frac{\|\mathsf{TT}\|+\|\mathsf{FT}\|+\|\mathsf{UT}\|}{\|\mathsf{TT}\|+\|\mathsf{FT}\|+\|\mathsf{UT}\|+\|\mathsf{TF}\|+\|\mathsf{FF}\|+\|\mathsf{UF}\|}
\end{equation*}		

The probability of \textit{$q$, known $p$}, on the other hand, looks exactly the same as in CTP. Given that $p$ is manifestly true, all non-$\mathsf{T}$ states of $p$ are ignored and according to \ref{tfudef}, so are all $\mathsf{U}$ states of $q$:

\begin{equation*}
	[q]_p=\frac{\|\mathsf{TT}\|}{\|\mathsf{TT}\|+\|\mathsf{TF}\|}
\end{equation*}		

Inverting the roles of $p$ and $q$, we get a fundamental result of a $\mathsf{TFU}$ probability theory based on  $\ref{tfudef}$:

\begin{equation}\label{uncommute}
	[p][q]_p\neq [q][p]_q
\end{equation}	

Let $\overline{p}:=$\textit{"$p$ is decidable"}. If $p$ is decidable, then so is $\neg p$, and vice versa:

\begin{equation*}
	\overline{p}\iff\overline{\neg{p}}
\end{equation*}	

The $\mathsf{TFU}$ states can now be defined in classical terms as:

\begin{equation*}
	\begin{cases}
		p\textrm{ is }\mathsf{T}:= p\land \overline{p} \\
		p\textrm{ is }\mathsf{F}:= \neg p\land \overline{p} \\
		p\textrm{ is }\mathsf{U}:= \neg\overline{p} 
	\end{cases}
\end{equation*}	

\ref{tfudef} can now be written as a function of classical probability as:

\begin{equation*}
	[p]=\frac{|p\land\overline{p}|}{|\overline{p}|}
\end{equation*}

And therefore:

\begin{equation*}
	[p]=|p|_{\overline{p}}
\end{equation*}

This curious relationship between three $p$ is the essence of $\textsf{TFU}$ probability; all that remains is to identify the algebra in which to place it. The vector version given by CTP is a good starting point, in the interest of exploiting the non-commutativity of the linear operators for the purpose of reproducing \ref{uncommute}.

In this particular representation, state $\ket{S}$ is a direction that forms the proper angles with the complete states

\begin{figure}[H]
	\centering
	\includegraphics[width=0.7\linewidth]{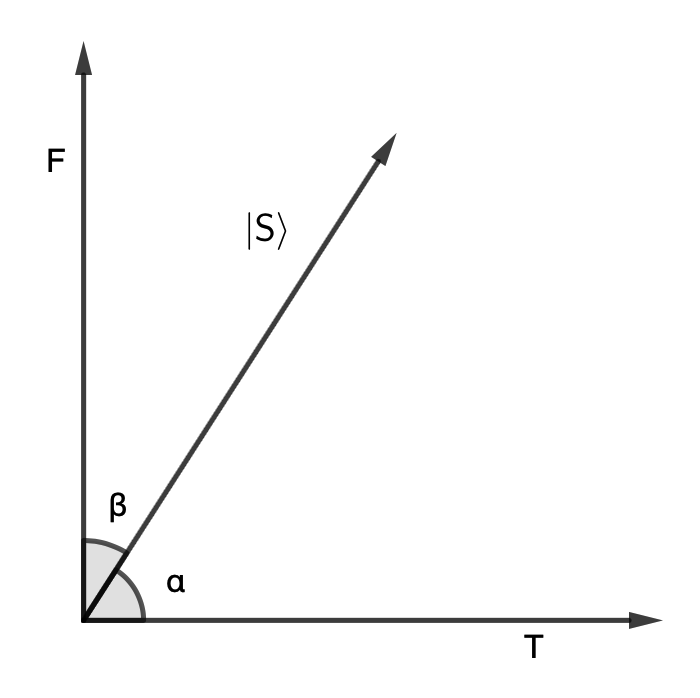}
	\caption[Didascalia]{}
	\label{fig:img2}
\end{figure}

In the simplest case of a single proposition, $\ket{S}$ forms angle $\alpha$ with the axis of state $\mathcal{T}$. The angle between $\ket{S}$ and the axis of state $\mathcal{F}$ is complementary to $\alpha$, and therefore this simple geometry reproduces the fundamental rule $|p|=1-|\neg p|$ in the form $cos^2\alpha=1-cos^2\beta$.

From a $\mathsf{TFU}$ perspective, the possible states are $3$ instead of $2$, considering an analogous geometrical representation in $\mathbb{R}^3$

\begin{figure}[H]
	\centering
	\includegraphics[width=0.7\linewidth]{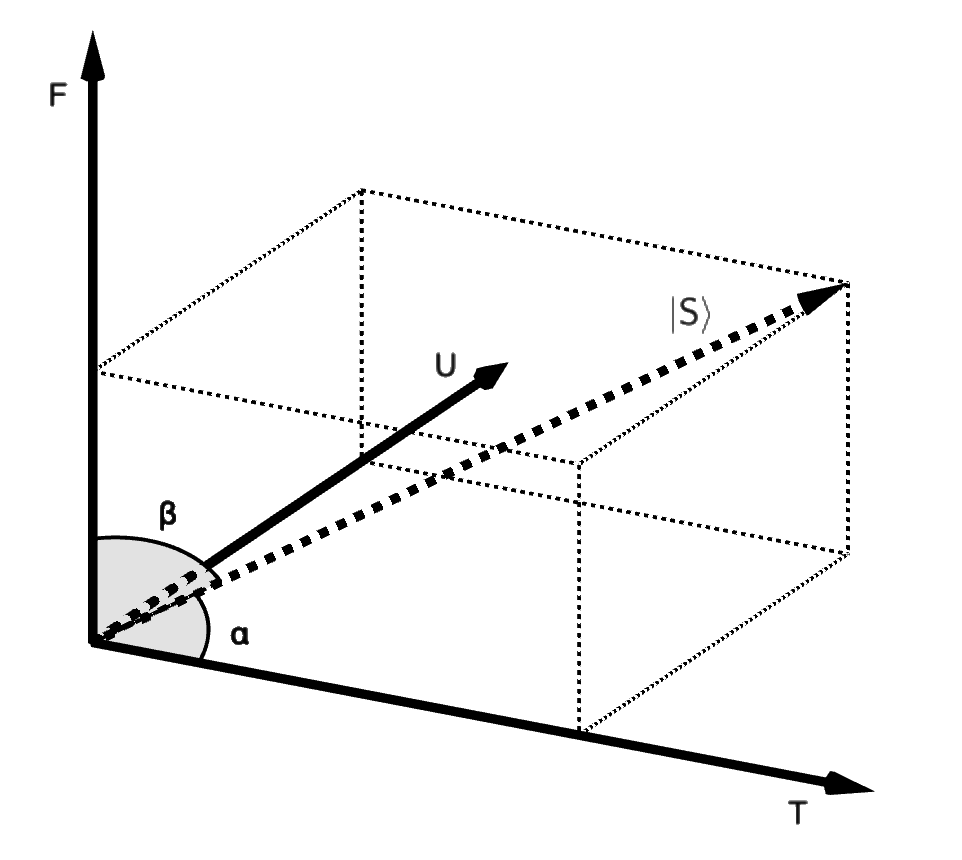}
	\caption[Didascalia]{}
	\label{fig:img3}
\end{figure}

This representation incorporates information on $\mathsf{TFU}$ states but angles $\alpha$ and $\beta$ are not complementary, therefore they lose the most characteristic feature, also valid in $\mathsf{TFU}$, of the vector representation given by CTP. The space suitable for representing a $\mathsf{TFU}$ version of $\ket{S}$ should instead present two complementary directions $\mathsf{T}$ and $\mathsf{F}$, and in any case provide a third degree of freedom that incorporates information related to state $\mathsf{U}$.

The extension being sought, therefore, is not from $\mathbb{R}^2$ to $\mathbb{R}^3$ but rather from $\mathbb{R}^2$ to $\mathbb{C}^2$.

Similarly, a problem involving $n$ propositions will be represented in space $\mathbb{C}^{2^n}=\mathbb{C}^2\otimes\mathbb{C}^2\otimes...$ and in such space a violation of \ref{uncommute} will be obtained by considering generic orthogonal projectors, thus, linear Hermitian operators with eigenvalues in $\{0,1\}$.

The $\mathsf{TFU}$ theory of probability therefore coincides with what is usually identified as \textit{quantum logic}.

\section{Conclusions}

Quantum mechanics is strange, precisely to the extent it violates Bell's inequalities and therefore classical probability, but it is not illogical and, in particular, it does not require a new logic nor violate any of the preconditions of classical logic.

The violation of Bell's inequalities makes it impossible to assign objective properties, but in a subtly relational form \cite{Rovelli}. A physical proposition, such as "the electron has spin-UP," for instance, is not "both true and false" nor even "neither true nor false". It may be considered objectively true or false ($\mathcal{T}$ o $\mathcal{F}$) but, in relation to the system of the observer, undecidable and this means that all such observer could do no more than adjust the weighting of the expected measures a priori with an algebra that violates CTP, thus running up against the usual paradoxes, which, from the right standpoint, avoiding the confusion of $\mathsf{TF}$ with $\mathcal{TF}$, are not paradoxical in any way.

The interpretation that emerges is clearly incompatible with attempts to \textit{"restore normality"} through the proliferation of objective states (many worlds, many minds). Alice has some properties $p$ that are undecidable for Bob; Bob's desire to resolve them by imagining the existence of a different Alice for every possible value of $p$ is understandable, but completely superfluous and unnecessarily costly.
QM is a relational theory that expresses the mechanics of a not completely knowable system, i.e., a system $S$ which, in relation to observer-system $O$, possesses some properties that are undecidable in $O$. Nothing, however, prevents that same property  $p$ of $S$, undecidable in $O$, from being decidable in a second system $S'$, which is also spatially separated from $S$. From the perspective of $O$, therefore, systems $S$ and $S'$ will be described as entangled. 

\subsection*{Conflicts of interest.}

The author declare no conflicts of interest.

\nocite{*}

\bibliography{Eng}

\end{document}